\documentclass[final,5p,times,twocolumn,number]{elsarticle}

\usepackage{amssymb}
\usepackage{lipsum}
\usepackage{amsmath}
\RequirePackage[colorlinks,citecolor=blue,urlcolor=blue,linkcolor=blue]{hyperref}
\usepackage{xcolor}
\RequirePackage{multicol}
\usepackage[skins,theorems,most]{tcolorbox}

\journal{Annals of Physics}

\begin{document}

\begin{frontmatter}

\title{Charged Hayward
black hole with a cosmological constant and  surrounded by quintessence and a cloud of strings}

\author[label1]{F. F. Nascimento}
\ead{fran.nice.fisica@gmail.com}
\author[label1]{V. B. Bezerra}
\ead{valdir@fisica.ufpb.br}

\author[label1]{J. M. Toledo}
\ead{jefferson.m.toledo@gmail.com}

\author[label3]{J. C. Rocha}
\ead{julio.rocha@servidor.uepb.edu.br}

\affiliation[label1]{organization={Departamento de Física, Universidade Federal da Paraíba, },
            addressline={Caixa Postal 5008}, 
            city={João Pessoa},
            postcode={58059-900}, 
            state={Paraíba},
            country={Brazil}}

\affiliation[label3]{organization={Departamento de Física, Universidade Estadual da Paraíba, },
            city={Campina Grande},
            postcode={58429-500}, 
            state={Paraíba},
            country={Brazil}}

\begin{abstract}
A family of exact solutions extending the Hayward black hole by incorporating multiple sources is obtained. The most comprehensive scenario describes a charged Hayward black hole with a cosmological constant, immersed in quintessence and accompanied by a cloud of strings. The study examines how the Kretschmann scalar varies with the parameters linked to the various sources and concludes with an analysis of the geodesics and the corresponding effective potential.
\end{abstract}


\end{frontmatter}

\section{Introduction}
\label{sec1}

A regular black hole is a solution of general relativity that, unlike classical static black holes, Schwarzschild (1916) \cite{schwarzschild1916uber} and Reiss-Nordström (1916) \cite{reissner1916eigengravitation, nordstrom1918een}, or stationary black holes, Kerr (1963) \cite{kerr1963gravitational} and Kerr-Newman (1965) \cite{newman1965metric}, does not have a physical singularity at its center. Therefore, a regular black hole is a compact object with an event horizon and without a singularity \cite{neves2017relatividade}. Regular solutions arise when models of matter are incorporated as sources, where the presence of a magnetic charge from the coupling between nonlinear electrodynamics and general relativity is considered \cite{ayon2000bardeen}. 

Among the first attempts to construct regular solutions, the work of Bardeen (1968) \cite{bardeen1968non} stands out, proposing a stable metric with an effective mass function that approaches Schwarzschild at large distances, but remains regular in the central region. Subsequently, Hayward (2006) \cite{hayward2006formation} developed a broader family of regular solutions, including models with “de Sitter core” characteristics in the interior, which preserve regularity and maintain the properties of possible horizons.  And recently, Frolov's model (2016) \cite{frolov2016notes}, which can basically be thought of as an extension of Hayward's spacetime with electric charge.

These approaches have established a conceptual framework for understanding how gravity can coexist with physical regularity under extreme conditions, paving the way for investigations into the nature of singularities, the stability of these solutions, and their possible observables.

In his 2003 article, V. V. Kiselev \cite{kiselev2003quintessence} proposed an exact solution in general relativity that describes a black hole immersed in a cosmological fluid with a linear equation of state given by $p=\omega \rho$. Although the main focus is on the description of black holes surrounded by an external energy with anisotropic properties, the construction can be interpreted as a link between the geometry of the space-time of a black hole and a cosmic energy field whose pressure is related to density by a parameter $\omega$. 

Thus, Kiselev's solution models a black hole immersed in a cosmological fluid with certain equation of state properties. The equation of state parameter $\omega$ allows modeling different energy regimes, such as radiation ($\omega=-1/3$), matter ($\omega\approx 0$), quintessence ($-1<\omega<-1/3$), and even phantom energy scenarios ($\omega<-1$).

Among the possible contributions of this model is the fact that it can offer a tool for studying how the presence of background energy affects local geometric properties near the horizon, as well as possible cosmological horizons induced by external energy, maintaining, in some cases, the regularity of the solution. Another fact concerns contributing to discussions about the nature of quintessence, the evolution of the universe on large scales, and the possible observable signatures of external fields in the geometry of black holes.

The cosmological constant $\Lambda$ appears in Einstein's field equations as a source of vacuum energy that acts as a uniform negative pressure throughout space. In cosmology, $\Lambda$ is strongly associated with dark energy, responsible for the acceleration of the expansion of the universe on a large scale \cite{riess1998observational,riess1999bvri,perlmutter1999measurements, ade2014planck}.
In regular black hole models that combine $\Lambda$ with other energy sources (quintessence, electric charge, cloud of strings), it can lead to new interesting classes of solutions with preserved or unpreserved regularity properties.

In 1978, Patrício S. Letelier presented an exact solution to Einstein's equations describing black holes surrounded by a cloud of strings \cite{letelier1979clouds}. This work inserts a specific anisotropic source into Einstein's set of equations, allowing us to study how a distribution of radial strings alters the gravitational field around compact objects. 

Among the possible motivations for studying cloud of strings is the desire to understand how extensive energy sources with anisotropic properties impact the structure of horizons, curvatures near black holes, and the topology of space-time in regions with strong gravity.

By investigating solutions that incorporate $\Lambda$ together with other energy sources, namely quintessence, electric charge, and cloud of strings, insights are gained into the interaction between gravity, cosmic expansion, and the nature of energy \cite{ghosh2015nonsingular, Ling_2023, rodrigues2022bardeenclounds, rodrigues2018bardeen, santos2024regular, nascimento2024somebardeen, nascimento2023some,lemos2011regular,nascimento2024black}.  Thus, in this work, in addition to deriving a general solution for the Hayward black hole with multiple sources, we will analyze the presence or absence of singularities in the general solution as well as in possible particular solutions based on the calculation of the Kretschmann scalar. We will also analyze the effective potential of the geodesic motion of massive and massless particles in the vicinity of the black hole with multiple sources.

The structure of this paper is as follows. In Sec. \ref{sec2} we revisit Hayward's solution, discussing some of its properties. In Sec. \ref{sec3} we obtain the solution describing a charged Hayward black hole with a cosmological constant, immersed in quintessence and accompanied by a cloud of strings. In Sec. \ref{sec4} we determine the Kretschmann scalar and examine the results for the general solution as well as for particular instances. In Sec. \ref{sec6}, we study the geodesics of the black hole. Our conclusions and perspectives can be found in Sec. \ref{sec7}.

\section{Hayward Black Hole}
\label{sec2}

Consider the metric below with spherical symmetry and describe a static black hole:
\begin{equation}
ds^2=f(r) dt^2-\frac{1}{f(r)}dr^2-r^2 d\theta^2-r^2\sin^2\theta d\phi^2,
\label{eq:1.03}
\end{equation}
such expression describes the Schwarzschild solution when the mass of the black hole is constant and $f(r)$ is given by
\begin{equation}
f(r)=1-\frac{2m}{r},
\label{eq:1.04}
\end{equation}
while describing the Hayward solution when the mass of the black hole is variable and $f(r)$ is given by:
\begin{equation}
f(r)=1-\frac{2M(r)}{r},
\label{eq:1.05}
\end{equation}
where
\begin{equation}
M(r)=\frac{mr^3}{r^3+2l^2m},
\label{eq:1.06}
\end{equation}
where $m$ is interpreted as a mass parameter, while $l$ is the Hayward parameter, whose value, in principle, will be restricted
to the range $0\leq l<\infty$. 

It is worth noting that Hayward, in his original publication, treated this parameter as being on the order of the Planck length. Subsequently linked it to a magnetic charge via the given definition \cite{bronnikov2001regular,molina2021thermodynamics,fan2016construction,bronnikov2017comment,toshmatov2018comment}

\begin{eqnarray}\label{magnetic charge}
q_m=\frac{\sqrt[3]{r_s^2l}}{2}=\frac{\sqrt[3]{4m^2l}}{2}.
\end{eqnarray}

\noindent It follows that the classical Schwarzschild radius is given by $r_s=2GM/c^2=2m$, where $m$ denotes the geometric mass and $M$ represents the system’s gravitational energy–mass. Consequently, the condition $q_m>0$ corresponds to a black-hole solution with $l>0$. 

From Eq. (\ref{eq:1.06}), one can see that as $r\rightarrow \infty$, the mass function satisfies $M(r)\rightarrow m$. Therefore, far from the black hole, the Hayward solution approaches the Schwarzschild solution.

\noindent For $r\rightarrow 0$, 
\begin{equation}
M(r) \approx \frac{r^3}{2l^2},
\label{eq:1.013}
\end{equation}
\noindent and 
\begin{equation}
f(r)\approx 1-Cr^2.
\label{eq:1.014}
\end{equation}

\noindent If $C=\frac{1}{l^2}>0$ is a positive constant, the metric function $f(r)$ given by Eq. (\ref{eq:1.05}) and (\ref{eq:1.06}) describes a space-time that is similar to the de Sitter space. Hence, the inner region of the black hole behaves like a de Sitter core \cite{hayward2006formation}.

There are broad strategies for diagnosing singularities. Among them is the analysis of invariants built from the curvature tensor, such as those derived from $R_{\mu\nu}$ or $R_{\mu\nu\sigma\rho}$, to determine whether a spacetime is singular.
The quantity that emerges enables a straightforward local assessment of whether the manifold under study contains singularities. Among the candidates, we focus on the Kretschmann scalar, denoted by $K$, defined as $K = R_{\mu\nu\sigma\rho}R^{\mu\nu\sigma\rho}$.

For the Hayward metric, we have the following Kretschmann scalar:

 \begin{equation}
K=\frac{48 m^2 \left[32 l^8 m^4-16 l^6 m^3 r^3+72 l^4 m^2 r^6-8 l^2 m r^9+r^{12}\right]}{\left(2 l^2 m+r^3\right)^6}.
\label{eq:1.011}
\end{equation}

\noindent In the limit $r \rightarrow 0$ and $r \rightarrow \infty$, we get
 
\begin{equation}
\lim_{r\rightarrow 0}K=\frac{24}{l^4}.
\label{eq:1.0011}
\end{equation}

\begin{equation}
\lim_{r\rightarrow \infty} K=0.
\label{eq:1.0012}
\end{equation}

Note that for a Schwarzschild black hole, the Kretschmann scalar diverges at $r = 0$, whereas for the Hayward black hole it remains finite as $r\rightarrow0$. As a result, the Hayward black hole, analyzed through the Kretschmann scalar, does not show a curvature singularity at the origin.

\subsection{Nonlinear magnetic monopole {$q_m$}}

In Hayward's solution, the source comes from nonlinear electrodynamics coupled to general relativity, an idea expressed by the action \cite{fan2016construction,molina2021thermodynamics,bronnikov2001regular,bronnikov2017comment,toshmatov2018comment}

\begin{eqnarray}\label{eq:1.015}
\begin{aligned}
S=\frac{1}{2\kappa^{2}}\int{d^{4}x\sqrt{-g}R}+\int{d^{4}x\sqrt{-g}L(F)}.
\end{aligned}
\end{eqnarray}

\noindent where $R$ is the scalar curvature and $L(F)$ is a nonlinear Lagrangian density, given by:

\begin{eqnarray}\label{eq:1.016}
L(F)=\frac{6 \left(2l^2F\right)^{3/2}}{\kappa ^2 l^2\left[1+\left(2l^2F\right)^{3/4}\right]^2},
\end{eqnarray}

\noindent which represents the particular source of nonlinear electrodynamics used to derive the Hayward black hole. In Eq. (\ref{eq:1.016}), the Lagrangian is a nonlinear function of the electromagnetic scalar $F=F^{\mu\nu}F_{\mu\nu}=2q_m^2/r^4$, where $F_{\mu\nu}$ is the Maxwell-Faraday tensor.

\section{Charged Hayward black hole with a cosmological
constant and surrounded by quintessence and a
cloud of strings.}
\label{sec3}

In this part, we derive the solution for a static, charged Hayward black hole in the presence of a cosmological constant, quintessence, and a cloud of strings ( Charged Hayward-AdS-Kiselev-Letelier black hole). This is achieved by solving Einstein’s equations in conjunction with a nonlinear electromagnetic field, incorporating the specified sources.

When the cosmological constant is included, the Einstein–Hilbert action becomes:

\begin{eqnarray}\label{actioEH}
S_{EH}=\frac{1}{2\kappa^{2}}\int{d^{4}x\,\sqrt{-g}\left(R+2\Lambda\right)}\,\mbox{.}
\end{eqnarray}

Regarding the portion of the action that arises from other sources and from the coupling with the nonlinear electromagnetic field, we denote this by $S_{CNLS}$. In this notation, $CNL$ denotes the coupling between the nonlinear electromagnetic field and the quintessence, which appears in the first term of $S_{CNLS}$ (as shown below). The symbols S represent the contributions associated with the electromagnetic field and the cloud of strings, corresponding to the second and third terms, respectively. Consequently, the action $S_{CNLS}$ can be expressed as:

\begin{eqnarray}\begin{aligned}
\label{actioM}
S_{CNLS}=&\int{d^{4}x\,\sqrt{-g}\,L(F)}\, + \int{d^{4}x\,\sqrt{-g}\,L_{E}}\\
&+\int{d^{4}x\,\sqrt{-g}\,L_{CS}}.
\end{aligned}
\end{eqnarray}

\noindent The nonlinear Lagrangian of electromagnetic theory  associated with Hayward space-time coupled to a quintessential fluid $L(F)$ \cite{hayward2006formation,bronnikov2001regular,molina2021thermodynamics,fan2016construction,rodrigues2022bardeen,bronnikov2017comment,toshmatov2018comment,kiselev2003quintessence} is described by, 

\begin{eqnarray}\label{L(F)}
L(F)=\frac{6 \left(2l^2F\right)^{3/2}}{\kappa ^2 l^2\left[1+\left(2l^2F\right)^{3/4}\right]^2}-\frac{6 \alpha  \omega  \left(\frac{F}{2 q_m^2}\right)^{\frac{3 (\omega +1)}{4}}}{\kappa ^2}.
\end{eqnarray}

\noindent In Eq. (\ref{L(F)}), the Lagrangian is a nonlinear function of the electromagnetic scalar $F=F^{\mu\nu}F_{\mu\nu}$, where $F_{\mu\nu}$ is the Maxwell-Faraday tensor.

In a spherically symmetric space-time with only magnetic charge, the sole nonzero component of $F_{\mu\nu}$ is 
\cite{bronnikov2001regular}

\begin{eqnarray}\label{F23}
F_{23}=q_m \sin{\theta},
\end{eqnarray}

\noindent with scalar $F$ given by

\begin{eqnarray}\label{F}
F=2F_{23}F^{23}=\frac{2q_m^2}{r^4},
\end{eqnarray}

\noindent where $q_m$ is the magnetic charge and $l$ is the Hayward parameter, both related by Eq. (\ref{magnetic charge}).

Thus, substituting Eq. (\ref{magnetic charge}) into Eq. (\ref{F}), we obtain:

\begin{eqnarray}\label{F(r_s)}
F=\frac{\sqrt[3]{r_s^4l^2}}{2r^4}=\frac{\sqrt[3]{2} \left(l m^2\right)^{2/3}}{r^4}.
\end{eqnarray}

\noindent For the electromagnetic field, the Lagrangian $L_E$ is given by \cite{d2022introducing}:
\begin{eqnarray}\label{L_E}
L_{E}=-\frac{1}{4}F_{\alpha\beta}F^{\alpha\beta},
\end{eqnarray}
whose non-zero components of the Maxwell tensor for a massive, static, charged and spherically symmetric object are:
\begin{eqnarray}\label{F01}
F_{01}=F_{10}=-F^{01}=E_{r}=\frac{Q}{r^2}.
\end{eqnarray}

For stringlike objects, the Nambu-Goto action is expressed as \cite{letelier1979clouds}

\begin{eqnarray}\label{S_{CS}}
S_{CS}=\int{d^{4}x\,\sqrt{-g}\,L_{CS}}\,=\int (-\gamma)^{1/2}\mathcal{M}{d\lambda^{0}d\lambda^{1}}.
\end{eqnarray}

The Lagrangian that describes the cloud of strings is $L_{CS}$ \cite{letelier1979clouds}:
\begin{eqnarray}\label{L_{CS}}
L_{CS} =\mathcal{M}\sqrt{-\gamma}= \mathcal{M}\left(-\frac{1}{2}\Sigma^{\mu\nu}\Sigma_{\mu\nu}\right)^{1/2}.
\end{eqnarray}

Therefore, the total action describing the dynamics of the charged Hayward black hole in the presence of a cosmological constant, enveloped by quintessence and a cloud of strings is given by
\begin{eqnarray}\label{total action}
\begin{aligned}
S=&S_{EH}+S_{CNLS}\\=&\frac{1}{2\kappa^{2}}\int{d^{4}x\sqrt{-g}\left(R+2\Lambda\right)}+\int{d^{4}x\sqrt{-g}L(F)}\\
+&\int{d^{4}x\sqrt{-g}L_{E}}+\int{d^{4}x\sqrt{-g}L_{CS}}.
\end{aligned}
\end{eqnarray}
Varying the action (\ref{total action}) with respect to the metric we find
\begin{eqnarray}\label{equationfield}
R_{\mu\nu}-\frac{1}{2}g_{\mu\nu}R -\Lambda g_{\mu\nu} = \kappa^{2}(T_{\mu\nu}^{CNL} + T_{\mu\nu}^{E} + T_{\mu\nu}^{CS}\,)\mbox{,}
\end{eqnarray}

\noindent accompanied by energy-momentum tensors described by \cite{bronnikov2001regular,d2022introducing,letelier1979clouds}:

\begin{eqnarray}\label{2.19}
T_{\mu\nu}^{CNL} = \frac{1}{2}g_{\mu\nu}L(F)-2\frac{\partial L}{\partial F}F_{\mu}^{\;\alpha}F_{\nu\alpha},
\end{eqnarray}

\begin{eqnarray}\label{2.21}
T_{\mu\nu}^{E} = \frac{1}{\kappa^2}\left(-2g^{\alpha \beta}F_{\mu \alpha}F_{\nu \beta}+\frac{1}{2}g_{\mu \nu}F_{\alpha \beta}F^{\alpha \beta}\right),
\end{eqnarray}

\begin{eqnarray}\label{2.22}
T_{\mu\nu}^{CS} = \frac{\rho\Sigma_{\mu}^{\;\beta}\Sigma_{\beta\nu}}{\kappa^2(-\gamma)^{1/2}}.
\end{eqnarray}.

\noindent The nonvanishing components of these tensors, specifically the $(0,0)$ and $(2,2)$ components, are given, respectively, by:

\begin{equation}\label{T00BK}
\begin{aligned}
T_{00}^{CNL} =\frac{12 l^2 m^2 f(r)}{\kappa ^2 \left(2 l^2 m+r^3\right)^2}-\frac{3 \alpha  \omega  f(r) r^{-3 \omega -3}}{\kappa ^2}.
\end{aligned}
\end{equation}

\begin{equation}\label{T22HK}
\begin{aligned}
T_{22}^{CNL} =\frac{24 l^2 m^2 r^2 \left(r^3-l^2 m\right)}{\kappa ^2 \left(2 l^2 m+r^3\right)^3}-\frac{3 \alpha  \omega  (3 \omega +1) r^{-3 \omega -1}}{2 \kappa ^2}.
\end{aligned}
\end{equation}

\begin{eqnarray}\label{T00E}
\begin{aligned}
T_{00}^{E} = \frac{f(r)}{\kappa^2}\frac{Q^2}{r^4},\;\;\;\;\;\;T_{22}^{E} =\frac{1}{\kappa^2}\frac{Q^2}{r^2}.
\end{aligned}
\end{eqnarray}

\begin{eqnarray}\label{T00CS}
\begin{aligned}
T_{00}^{CS} =\frac{f(r)}{\kappa^2}\frac{a}{r^2},\;\;\;\;\;\;T_{22}^{CS} =0.
\end{aligned}
\end{eqnarray}

\noindent Accounting for these results of the energy-momentum tensors, Eq. (\ref{equationfield}) simplifies to the subsequent differential equations

\begin{equation}
-\frac{a}{r^2}-\frac{f'(r)}{r}-\frac{f(r)}{r^2}+\frac{1}{r^2}-\Lambda -\frac{12 l^2 m^2}{\left(2 l^2 m+r^3\right)^2}-\frac{Q^2}{r^4}+3 \alpha  \omega  r^{-3 \omega -3}=0,
\label{edo1}
\end{equation}

\begin{equation}
\begin{aligned}
&\frac{1}{2} r^2 f''(r)+r f'(r)-\frac{24 l^2 m^2 r^2 \left(r^3-l^2 m\right)}{\left(2 l^2 m+r^3\right)^3}-\frac{Q^2}{r^2}+\Lambda  r^2
\\
&+\frac{3}{2} \alpha  \omega  (3 \omega +1) r^{-3 \omega -1}=0.
\end{aligned}
\label{edo2}
\end{equation}

\noindent Multiplying Eq. (\ref{edo1}) by $r^2$ and Eq. (\ref{edo2}) by $2$, adding the results, we get:

\begin{equation}
\begin{aligned}
&1-a-f(r)+r f'(r)+r^2 f''(r)\\
&-\frac{12 l^2 m^2 r^2}{\left(2 l^2 m+r^3\right)^2}-\frac{48 l^2 m^2 r^2 \left(r^3-l^2 m\right)}{\left(2 l^2 m+r^3\right)^3}\\
&-\frac{3 Q^2}{r^2}+3 \alpha  \omega  (3 \omega +2) r^{-3 \omega -1}+\Lambda  r^2=0.
\end{aligned}
\label{edo3}
\end{equation}

\noindent Solving the differential equation above we have

\begin{equation}
f(r)=1-a+\frac{4 l^2 m^2}{2 l^2 m r+r^4}+\frac{Q^2}{r^2}-\alpha  r^{-3 \omega -1}-\frac{\Lambda  r^2}{3}+\frac{C_1}{r}+C_2 r.
\label{solution}
\end{equation}

\noindent Let $C_1$ and $C_2$ be arbitrary constants whose values are determined by applying certain conditions that the physical system must satisfy. We should adopt $C_1=-2m$ such that, when combined with the third term of Eq. (\ref{solution}), it recovers Hayward's solution when all additional sources are absent.

Note that our analysis treats quintessence as one of the sources, which fixes the equation of state parameter to $\omega_q=-2/3$. Consequently, the fifth and eighth terms in Eq. (\ref{solution}) can be merged into a single effective constant.
With these results in hand, the function $f(r)$
can be expressed in a simplified form that omits the last two terms. Employing this streamlined expression, the metric describing a charged Hayward black hole with a cosmological constant, embedded in quintessence and a cloud of strings, within a nonlinear electrodynamics framework coupled to Einstein gravity, is given by:

\begin{eqnarray}\label{eq:1.31}
\begin{aligned}
&ds^{2} =\\
&+\left(1-a-\frac{2 m r^2}{r^3+2 l^2 m}-\frac{\alpha}{r^{3 \omega _q+1}}+\frac{Q^2}{r^2}-\frac{1}{3}\Lambda  r^2\right)dt^{2}\\
&-\left(1-a-\frac{2 m r^2}{r^3+2 l^2 m}-\frac{\alpha}{r^{3 \omega _q+1}}+\frac{Q^2}{r^2}-\frac{1}{3}\Lambda  r^2\right)^{-1}dr^{2}\\
&- r^{2}d\theta^{2} - r^{2}\sin^{2}\theta d\phi^{2}.
\end{aligned}
    \end{eqnarray}

\noindent In Table \ref{table_dif_ST}, we describe the spacetime that can be recovered from Eq. (\ref{eq:1.31}).

\begin{table}[h]
\caption{Space-time that can be recovered from the Eq. (\ref{eq:1.31})}
\begin{tabular}{lcclccclcccclcl}
\hline
&         BH PARAMETERS                  &     SPACE-TIME           &\\\hline
&          $a=\alpha=Q=\Lambda=0$                        &     Hayward              &\\
&         $a=\alpha=\Lambda=0$                            &     Hayward-Reissner-Nordström&\\
&          $\alpha=Q=\Lambda=0$                           &     Hayward-Letelier&\\
&         $a=Q=\Lambda=0$                          &     Hayward-Kiselev&\\
&         $a=\alpha=Q=0$                           &     Hayward-AdS&\\
&         $Q=\Lambda=0$                           &     Hayward-Kiselev-Letelier&\\
&          $Q=0$                           &     Hayward-AdS-Kiselev-Letelier&\\\hline
\end{tabular}
\label{table_dif_ST}
\end{table}


\section{Kretschmann Scalar Analysis}
\label{sec4}

To check for singularities in this space-time, a common approach is to evaluate the Kretschmann scalar. For the metric described by Eq. (\ref{eq:1.31}), the Kretschmann scalar is obtained as:

\begin{equation}
\begin{aligned}
K&=R_{\alpha\beta\mu\nu}R^{\alpha\beta\mu\nu}
\\
&=\frac{4 a^2}{r^4}+\frac{16 a m}{2 l^2 m r^2+r^5}-\frac{8 a Q^2}{r^6}+\frac{8 a \Lambda }{3 r^2}
\\
&+8 a \alpha r^{-3\omega_q-5}+4 \alpha \Lambda  \omega _q \left(3 \omega _q-1\right) r^{-3 \left(\omega_q+1\right)}
\\
&+\frac{48 m^2 \left[32 l^8 m^4-16 l^6 m^3 r^3+72 l^4 m^2 r^6-8 l^2 m r^9+r^{12}\right]}{\left(2 l^2 m+r^3\right)^6}
\\
&+\frac{8 \Lambda  \left[8 \Lambda  l^6 m^3+12 l^4 m^2 \left(4 m+\Lambda  r^3\right)\right]}{3 \left(2 l^2 m+r^3\right)^3}
\\
&+\frac{8 \Lambda  \left[6 l^2 m r^3 \left(\Lambda  r^3-2 m\right)+\Lambda  r^9\right]}{3 \left(2 l^2 m+r^3\right)^3}
\\
&+\frac{8 Q^2 \left[56 l^6 m^3 Q^2+84 l^4 m^2 Q^2 r^3\right]}{r^8 \left(2 l^2 m+r^3\right)^3}
\\
&+\frac{8 Q^2 \left[42 l^2 m r^6 \left(2 m r+Q^2\right)+r^9 \left(7 Q^2-12 m r\right)\right]}{r^8 \left(2 l^2 m+r^3\right)^3}
\\
&+3 \alpha^2 \left(27 \omega _q^4+54 \omega _q^3+51 \omega _q^2+20 \omega _q+4\right) r^{-6 \left(\omega _q+1\right)}
\\
&-12\alpha Q^2 \left(9\omega_q^2+13\omega_q+4\right) r^{-3\omega_q-7}
\\
&+\frac{24 \alpha m r^{-3 \left(\omega _q+1\right)} \left[4 l^4 m^2 \omega _q \left(3 \omega _q-1\right)\right]}{\left(2 l^2 m+r^3\right)^3}
\\
&+\frac{24 \alpha m r^{-3 \left(\omega _q+1\right)} \left[-2 l^2 m r^3 \left(21 \omega _q^2+23 \omega _q+4\right)\right]}{\left(2 l^2 m+r^3\right)^3}
\\
&+\frac{24 \alpha m r^{-3 \left(\omega _q+1\right)} \left[r^6 \left(3 \omega _q^2+5 \omega _q+2\right)\right]}{\left(2 l^2 m+r^3\right)^3}.
\label{eq:1.32}
\end{aligned}
\end{equation}

\begin{figure*}[h]
\centering
\begin{minipage}[!]{0.43\linewidth}
\includegraphics[scale=0.6]{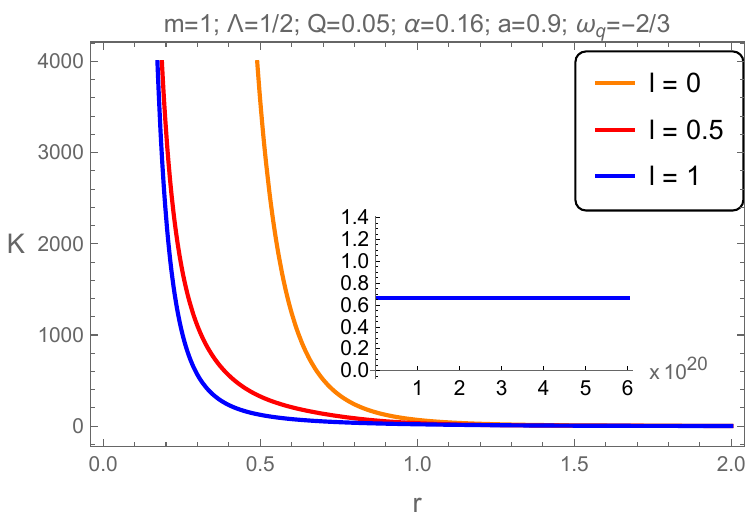}
\vspace{2ex}
      \end{minipage}
\begin{minipage}[!]{0.43\linewidth}
\includegraphics[scale=0.6]{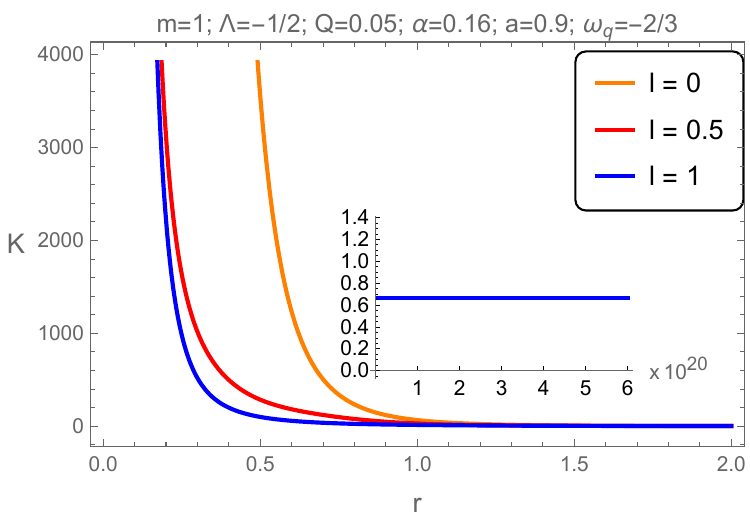}
\vspace{2ex}
     \end{minipage}
\begin{minipage}[!]{0.43\linewidth}
\includegraphics[scale=0.6]{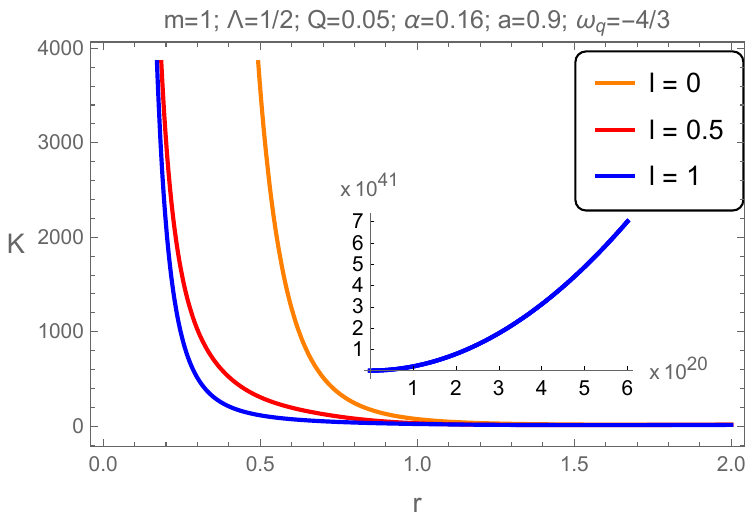}
     \end{minipage}
\begin{minipage}[!]{0.43\linewidth}
\includegraphics[scale=0.6]{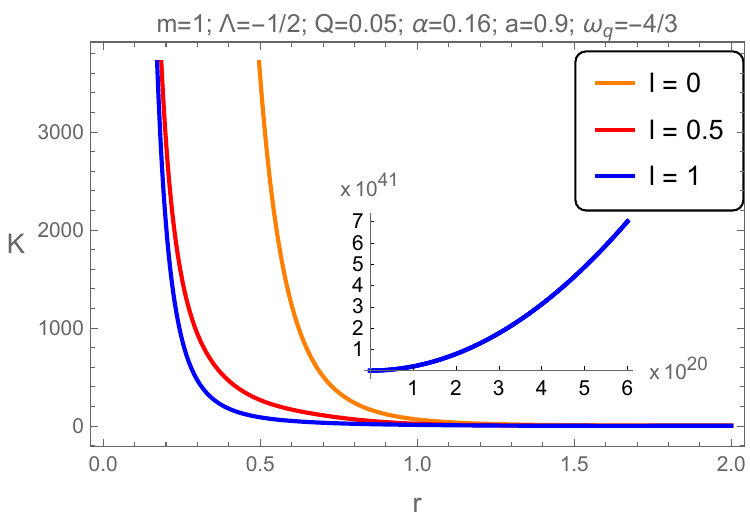}
     \end{minipage}
\caption{Kretschmann scalar referring to the charged Hayward  black hole with a cosmological constant and surrounded by quintessence and a cloud of strings, for different values of the parameters $l$, $\Lambda$, and $\omega_q$.} 
\label{im1}
\end{figure*}

\noindent Let us now determine the limit of the Kretschmann scalar with $r\rightarrow 0$ and $r\rightarrow \infty$. For $\omega_q=-2/3$ (quintessence energy regime), we get the following limits:
 
\begin{equation}
\lim_{r\rightarrow 0}K=\infty.
\label{eq:1.39}
\end{equation}

\begin{equation}
\lim_{r\rightarrow \infty}K=\frac{8 \Lambda ^2}{3}.
\label{eq:1.40}
\end{equation}

\noindent Furthermore, we examine the energy regime known as phantom, where $\omega_q=-4/3$. Consequently, we obtain the following limits

\begin{equation}
\lim_{r\rightarrow 0}K=\infty.
\label{eq:1.35}
\end{equation}

\begin{equation}
\lim_{r\rightarrow \infty}K=\infty.
\label{eq:1.36}
\end{equation}

In summary, the curvature remains finite as $r\rightarrow \infty$ and $\omega_q=-2/3$, but it diverges at the origin. For $\omega_q=-4/3$, the solution is singular in both limits, as shown in Fig. \ref{im1}.

\subsection{Hayward black hole}

By making $\alpha=0$, $\Lambda=0$, $Q=0$ and $a=0$ in Eq. (\ref{eq:1.31}), we obtain the Hayward black hole given by:

\begin{equation}
 \begin{aligned}
ds^2=&\left(1-\frac{2 m r^2}{r^3+2 l^2 m}\right)dt^2\\
&-\left(1-\frac{2 m r^2}{r^3+2 l^2 m}\right)^{-1}dr^2-r^2 d\Omega^2,
\label{eq:1.41}
\end{aligned} 
\end{equation}

\noindent whose Kretschmann scalar is

\begin{equation}
K=\frac{48 m^2 \left[32 l^8 m^4-16 l^6 m^3 r^3+72 l^4 m^2 r^6-8 l^2 m r^9+r^{12}\right]}{\left(2 l^2 m+r^3\right)^6},
\label{eq:1.42}
\end{equation}

\noindent where
 
\begin{equation}
\lim_{r\rightarrow 0}K=\frac{24}{l^4}\,\,\,\mbox{and}\,\,\,\lim_{r\rightarrow \infty}K=0.
\label{eq:1.43}
\end{equation}

\begin{figure*}[h]
\centering
\includegraphics[scale=0.8]{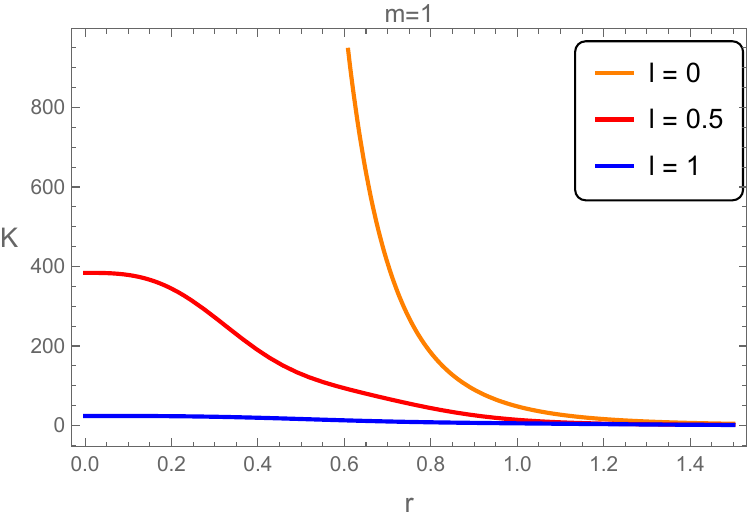}
\caption{Kretschmann scalar referring to the Hayward black hole for different values of $l$.} 
\label{im2}
\end{figure*}

\begin{figure*}[h]
\centering
\begin{minipage}[!]{0.45\linewidth}
\includegraphics[scale=0.6]{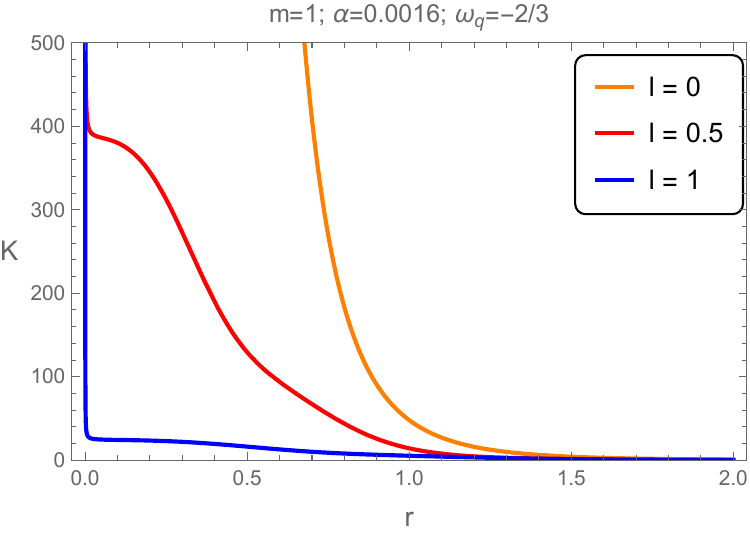}
\vspace{2ex}
      \end{minipage}
\begin{minipage}[!]{0.45\linewidth}
\includegraphics[scale=0.6]{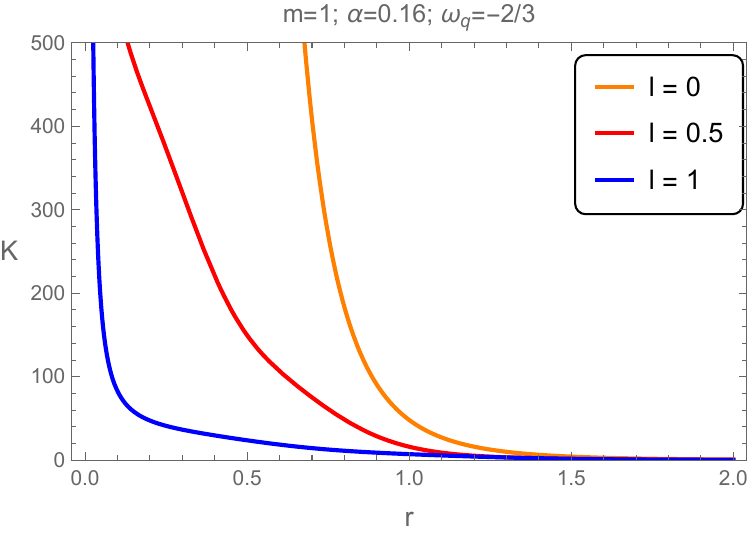}
\vspace{2ex}
     \end{minipage}
\begin{minipage}[!]{0.45\linewidth}
\includegraphics[scale=0.6]{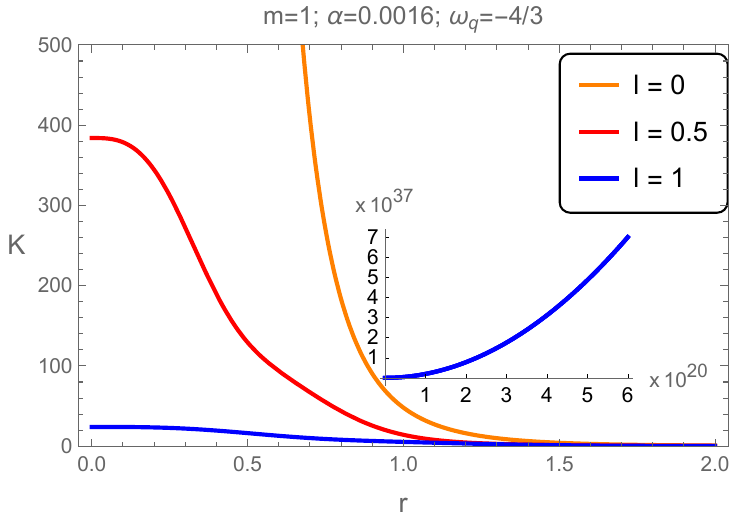}
     \end{minipage}
\begin{minipage}[!]{0.45\linewidth}
\includegraphics[scale=0.6]{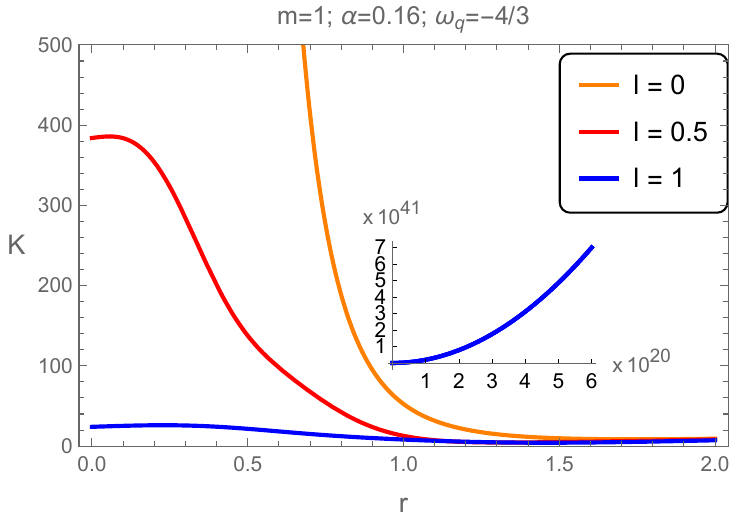}
     \end{minipage}
\caption{Kretschmann scalar referring to the Hayward-Kiselev black hole for different values of $l$, $\omega_q$ and $\alpha$.} 
\label{im3}
\end{figure*}

Thus, the analysis of the Kretschmann scalar in the limit of $r\rightarrow 0$ and $r\rightarrow \infty$ shows us, as expected, that the Hayward black hole is regular, that is, it has no singularity at the origin $r=0$ and is asymptotically flat. This behavior is described in Fig. \ref{im2}.

\subsection{Hayward black hole surrounded by quintessence}

By making $\Lambda=0$, $Q=0$ and $a=0$ in Eq. (\ref{eq:1.31}), we obtain the Hayward black hole surrounded by quintessence (Hayward-Kiselev black hole) given by:

\begin{equation}
 \begin{aligned}
ds^2=&\left(1-\frac{2 m r^2}{r^3+2 l^2 m}-\frac{\alpha}{r^{3 \omega _q+1}}\right)dt^2\\
&-\left(1-\frac{2 m r^2}{r^3+2 l^2 m}-\frac{\alpha}{r^{3 \omega _q+1}}\right)^{-1}dr^2-r^2 d\Omega^2,
\label{eq:1.45}
\end{aligned} 
\end{equation}

\noindent with the Kretschmann scalar given by

\begin{equation}
\begin{aligned}
K&=\frac{48 m^2 \left[32 l^8 m^4-16 l^6 m^3 r^3+72 l^4 m^2 r^6-8 l^2 m r^9+r^{12}\right]}{\left(2 l^2 m+r^3\right)^6}
\\
&+\frac{24 \alpha m r^{-3 \left(\omega _q+1\right)} \left[4 l^4 m^2 \omega _q \left(3 \omega _q-1\right)\right]}{\left(2 l^2 m+r^3\right)^3}
\\
&+\frac{24 \alpha m r^{-3 \left(\omega _q+1\right)} \left[-2 l^2 m r^3 \left(21 \omega _q^2+23 \omega _q+4\right)\right]}{\left(2 l^2 m+r^3\right)^3}
\\
&+\frac{24 \alpha m r^{-3 \left(\omega _q+1\right)} \left[r^6 \left(3 \omega _q^2+5 \omega _q+2\right)\right]}{\left(2 l^2 m+r^3\right)^3}
\\
&+3 \alpha^2 \left(27 \omega _q^4+54 \omega _q^3+51 \omega _q^2+20 \omega _q+4\right) r^{-6 \left(\omega _q+1\right)}.
\label{eq:1.46}
\end{aligned}
\end{equation}

\noindent For $\omega_q=-2/3$, quintessence energy regime, we get the following limits of the Kretschmann scalar:

\begin{equation}
\lim_{r\rightarrow 0}K=\infty\,\,\,\mbox{e}\,\,\,\lim_{r\rightarrow \infty}K=0.
\label{eq:1.51}
\end{equation}

\noindent For $\omega_q=-4/3$, phantom energy regime, we get the following limits:

\begin{equation}
\lim_{r\rightarrow 0}K=\frac{24}{l^4}\,\,\,\mbox{e}\,\,\,\lim_{r\rightarrow \infty}K=\infty.
\label{eq:1.53}
\end{equation}

Thus, the analysis of the Kretschmann scalar in the limit of $r\rightarrow 0$ and $r\rightarrow \infty$, whose behavior is described in Fig. \ref{im3}, shows us that:

\begin{itemize}
\item the Hayward black hole surrounded by quintessence is not regular when $\omega_q=-2/3$; it possesses a singularity at the origin $r=0$ but remains asymptotically flat.
\item The Hayward black hole with phantom is regular for $\omega_q=-4/3$, without singularity at the origin $r=0$; however, it is not asymptotically flat as $r\rightarrow \infty$.
\end{itemize}

\subsection{Hayward black hole with Cosmological Constant}

By making $\alpha=0$, $Q=0$ and $a=0$ in Eq. (\ref{eq:1.31}), we obtain the Hayward black hole with cosmological constant (Hayward AdS black hole), given by:

\begin{equation}
 \begin{aligned}
ds^2=&\left(1-\frac{2 m r^2}{r^3+2 l^2 m}-\frac{1}{3}\Lambda  r^2\right)dt^2\\
&-\left(1-\frac{2 m r^2}{r^3+2 l^2 m}-\frac{1}{3}\Lambda  r^2\right)^{-1}dr^2-r^2 d\Omega^2,
\label{eq:1.55}
\end{aligned} 
\end{equation}

\noindent where the Kretschmann scalar is

\begin{equation}
\begin{aligned}
K&=\frac{48 m^2 \left[32 l^8 m^4-16 l^6 m^3 r^3+72 l^4 m^2 r^6-8 l^2 m r^9+r^{12}\right]}{\left(2 l^2 m+r^3\right)^6}
\\
&+\frac{8 \Lambda  \left[8 \Lambda  l^6 m^3+12 l^4 m^2 \left(4 m+\Lambda  r^3\right)+6 l^2 m r^3 \left(\Lambda  r^3-2 m\right)+\Lambda  r^9\right]}{3 \left(2 l^2 m+r^3\right)^3},
\label{eq:1.56}
\end{aligned}
\end{equation}

 \noindent and
 
\begin{equation}
\lim_{r\rightarrow 0} K=\frac{8 \left(\Lambda  l^2+3\right)^2}{3 l^4}.
\label{eq:1.57}
\end{equation}

\begin{equation}
\lim_{r\rightarrow \infty} K=\frac{8\Lambda^2}{3}.
\label{eq:1.58}
\end{equation}

\begin{figure*}
\centering
\begin{minipage}[!]{0.46\linewidth}
\includegraphics[scale=0.6]{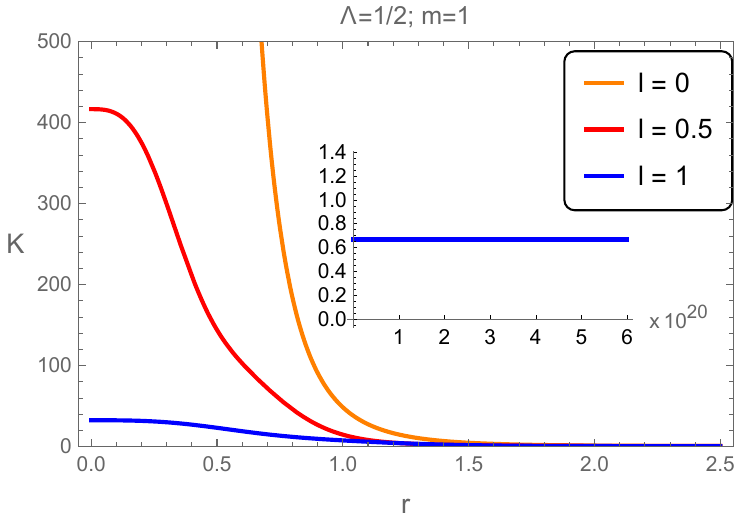}
      \end{minipage}
\begin{minipage}[!]{0.46\linewidth}
\includegraphics[scale=0.6]{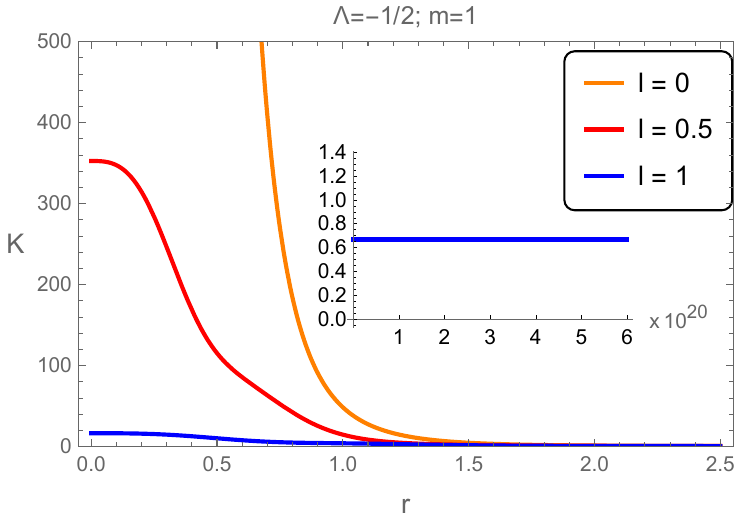}
     \end{minipage}
\caption{Kretschmann scalar referring to the Hayward black hole surrounded by quintessence
 for different values of $l$, $\omega_q$, and $\alpha$.} 
\label{im4}
\end{figure*}

Thus, we conclude that the inclusion of the cosmological constant to the Hayward black hole does not affect
the regularity of the metric as can also be seen in Fig. \ref{im4}.

\subsection{Hayward black hole with electromagnetic field}

By making $\alpha=0$, $\Lambda=0$ and $a=0$ in Eq. (\ref{eq:1.31}), we obtain the Hayward black hole with electromagnetic field (Hayward-Reissner-Nordström black hole), given by:

\begin{equation}
 \begin{aligned}
ds^2=&\left(1-\frac{2 m r^2}{r^3+2 l^2 m}+\frac{Q^2}{r^2}\right)dt^2\\
&-\left(1-\frac{2 m r^2}{r^3+2 l^2 m}+\frac{Q^2}{r^2}\right)^{-1}dr^2-r^2 d\Omega^2,
\label{eq:1.59}
\end{aligned} 
\end{equation}

\noindent whose Kretschmann scalar is
 
 \begin{equation}
\begin{aligned}
K&=\frac{48 m^2 \left[32 l^8 m^4-16 l^6 m^3 r^3+72 l^4 m^2 r^6-8 l^2 m r^9+r^{12}\right]}{\left(2 l^2 m+r^3\right)^6}
\\
 &+\frac{8 Q^2 \left[56 l^6 m^3 Q^2+84 l^4 m^2 Q^2 r^3\right]}{r^8 \left(2 l^2 m+r^3\right)^3}
 \\
 &+\frac{8 Q^2 \left[42 l^2 m r^6 \left(2 m r+Q^2\right)+r^9 \left(7 Q^2-12 m r\right)\right]}{r^8 \left(2 l^2 m+r^3\right)^3}.
 \label{eq:1.60}
\end{aligned}
\end{equation}

 \noindent and

\begin{equation}
\lim_{r\rightarrow 0}K=\infty\,\,\,\mbox{and}\,\,\,\lim_{r\rightarrow \infty}K=0.
\label{eq:1.61}
\end{equation}

\begin{figure*}
\centering
\begin{minipage}[!]{0.46\linewidth}
\includegraphics[scale=0.6]{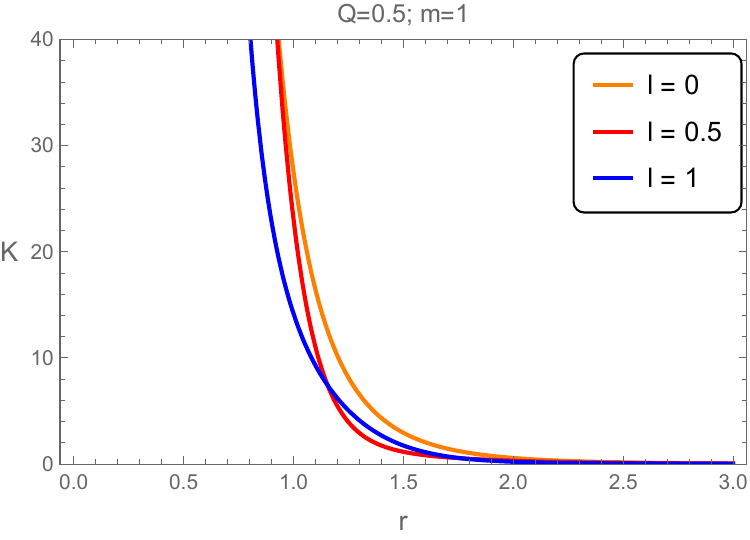}
      \end{minipage}
\begin{minipage}[!]{0.46\linewidth}
\includegraphics[scale=0.6]{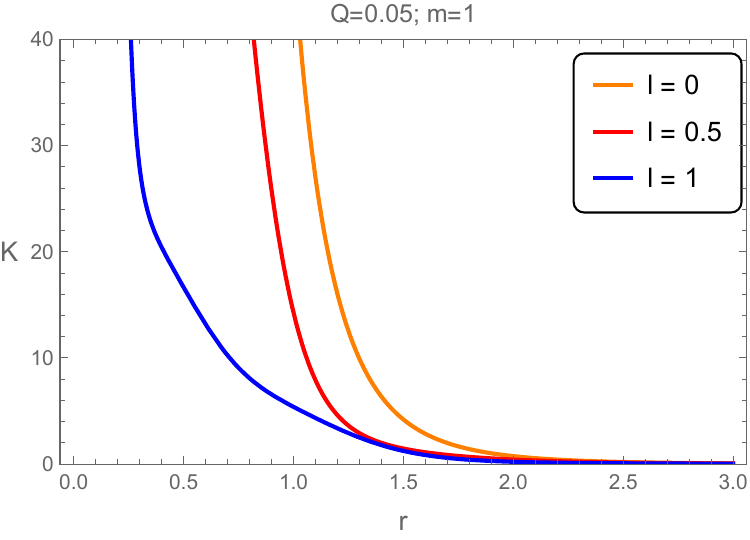}
     \end{minipage}
\caption{Kretschmann scalar referring to the Hayward black hole with electromagnetic field for different values of $l$ and $Q$.} 
\label{im5}
\end{figure*}

Thus, we conclude that the inclusion of the electric charge in Hayward's metric affects the regularity of the metric by making it singular at the origin. The metric remains asymptotically flat. This behavior can be seen in Fig. \ref{im5}.
\subsection{Hayward black hole with cloud of strings}

By making $\alpha=0$, $\Lambda=0$ and $Q=0$ in Eq. (\ref{eq:1.31}), we obtain the Hayward black hole with cloud of strings (Hayward-Letelier Black Hole) \cite{nascimento2023some}, given by:

\begin{equation}
 \begin{aligned}
ds^2=&\left(1-a-\frac{2 m r^2}{r^3+2 l^2 m}\right)dt^2\\
&-\left(1-a-\frac{2 m r^2}{r^3+2 l^2 m}\right)^{-1}dr^2-r^2 d\Omega^2,
\label{eq:1.63}
\end{aligned} 
\end{equation}

\noindent whose Kretschmann scalar is
 
 \begin{equation}
\begin{aligned}
K&=\frac{4 a^2}{r^4}+\frac{16 a m}{2 l^2 m r^2+r^5}
\\
&+\frac{48 m^2 \left[32 l^8 m^4-16 l^6 m^3 r^3+72 l^4 m^2 r^6-8 l^2 m r^9+r^{12}\right]}{\left(2 l^2 m+r^3\right)^6},
\label{eq:1.64}
\end{aligned}
\end{equation}

 \noindent and

 \begin{equation}
\lim_{r\rightarrow 0}K=\infty\,\,\,\mbox{and}\,\,\,\lim_{r\rightarrow \infty}K=0.
\label{eq:1.65}
\end{equation}

\begin{figure*}
\centering
\begin{minipage}[!]{0.46\linewidth}
\includegraphics[scale=0.6]{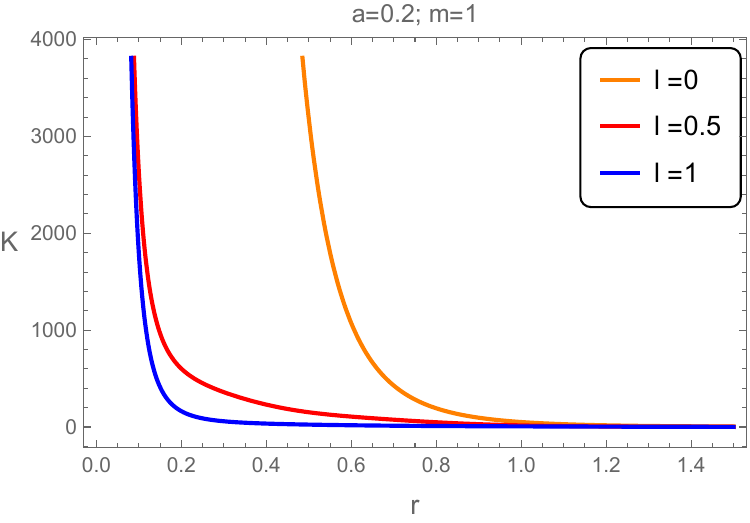}
      \end{minipage}
\begin{minipage}[!]{0.46\linewidth}
\includegraphics[scale=0.6]{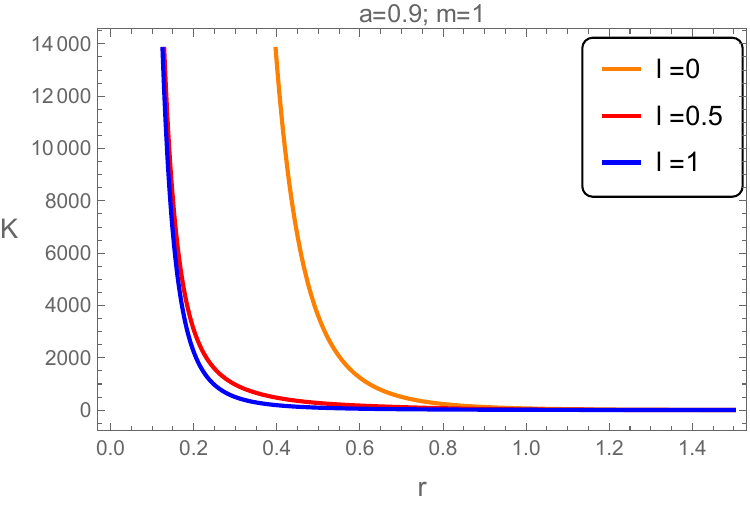}
     \end{minipage}
\caption{Kretschmann scalar referring to the Hayward black hole with cloud of strings for different values of $l$ and $a$.} 
\label{im6}
\end{figure*}

The cloud of strings surrounding the Hayward black hole destroys the regularity at the origin of the original solution. The spacetime remains asymptotically flat, as illustrated in Fig. \ref{im6}.

\subsection{Hayward black hole with cosmological constant and surrounded by quintessence}

From Eq. (\ref{eq:1.32}), it turns out that, for Hayward-Kiselev-AdS black hole ($a=0$ and $Q=0$), space-time is regular only for some values of $\omega_q$. For $\omega_q=-2/3$ (quintessence energy regime) the solution is singular for $r\rightarrow 0$ and has a regular curvature for $r\rightarrow \infty$. 

\begin{equation}
\lim_{r\rightarrow 0}K=\infty,
\label{eq:1.67}
\end{equation}

\begin{equation}
\lim_{r\rightarrow \infty}K=\frac{8 \Lambda ^2}{3}.
\label{eq:1.68}
\end{equation}

\noindent For $\omega_q=-4/3$ (phantom energy regime), the solution is regular for $r\rightarrow 0$ and has not a regular curvature for $r\rightarrow \infty$. 

\begin{equation}
\lim_{r\rightarrow 0}K=\frac{8 \left(\Lambda  l^2+3\right)^2}{3 l^4},
\label{eq:1.69}
\end{equation}

\begin{equation}
\lim_{r\rightarrow \infty}K=\infty.
\label{eq:1.70}
\end{equation}

\noindent For $\omega_q=-1$, we have the following limits: 

\begin{equation}
\lim_{r\rightarrow 0}K=\frac{8 \left(3 b l^2+\Lambda  l^2+3\right)^2}{3 l^4}.
\label{eq:1.74}
\end{equation}

\begin{equation}
\lim_{r\rightarrow \infty}K=\frac{8}{3} (3 b+\Lambda )^2.
\label{eq:1.74.1}
\end{equation}

\noindent For this value of $\omega_q$, the solution is curvature regular everywhere, since we have only a shift in the cosmological constant.

\begin{figure*}[h]
\centering
\begin{minipage}[!]{0.45\linewidth}
\includegraphics[scale=0.6]{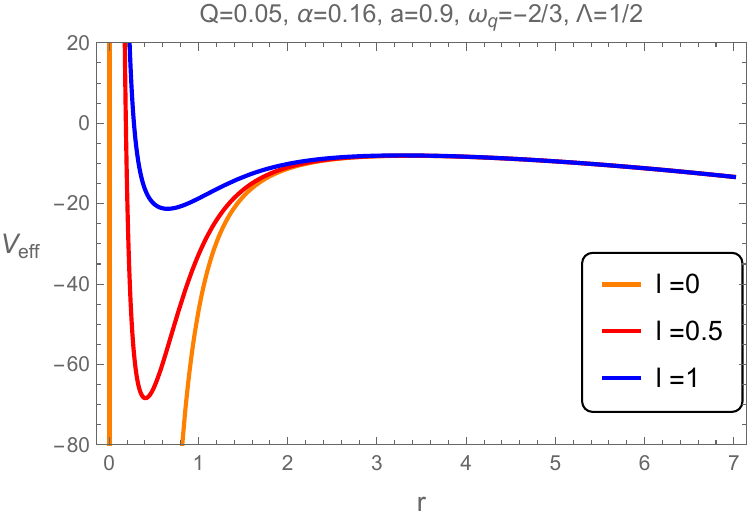}
\vspace{2ex}
      \end{minipage}
\begin{minipage}[!]{0.45\linewidth}
\includegraphics[scale=0.6]{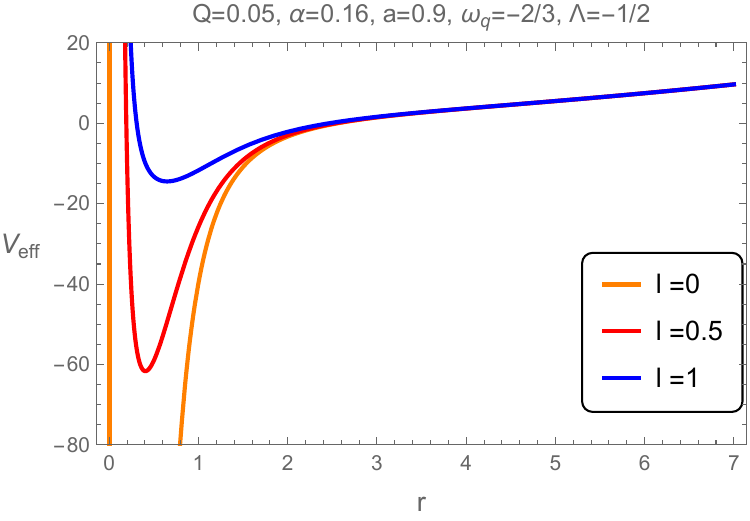}
\vspace{2ex}
     \end{minipage}
\begin{minipage}[!]{0.45\linewidth}
\includegraphics[scale=0.6]{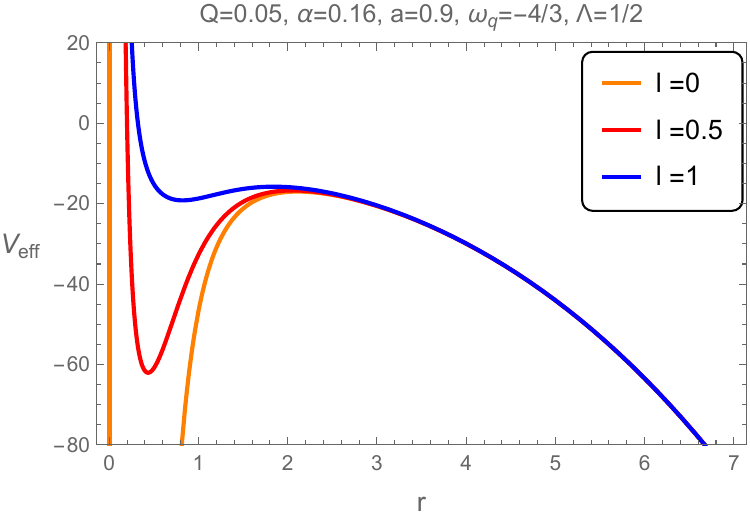}
     \end{minipage}
\begin{minipage}[!]{0.45\linewidth}
\includegraphics[scale=0.6]{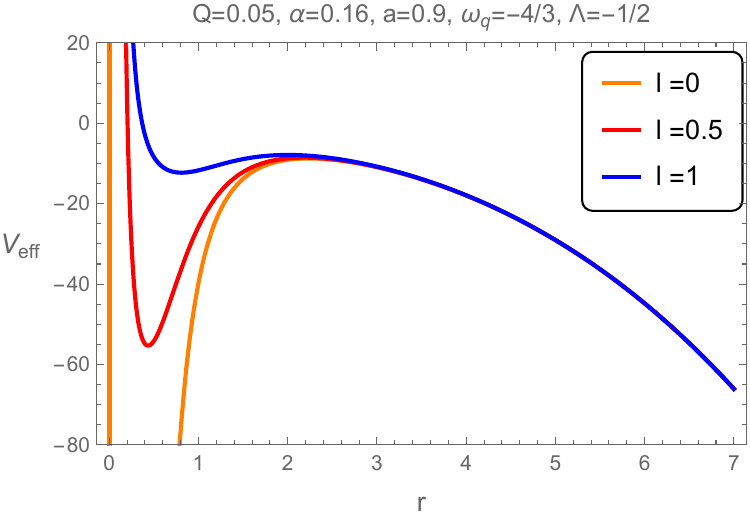}
     \end{minipage}
\caption{Effective potential for non-radial time-like geodesics ($L=1$ and $J^2=20$), for different values of $l$, $\omega_q$ and $\Lambda$.} 
\label{im7}
\end{figure*}

\section{Geodesics and effective potential}
\label{sec6}

The geodesic equations can be derived from the Lagrangian presented in the equation

\begin{equation*}
\mathcal{L}=\frac{1}{2}g_{\mu\nu}\frac{dx^\mu}{d\tau}\frac{dx^\nu}{d\tau}=\frac{1}{2} g_{\mu\nu}\dot{x}^\mu\dot{x}^\nu,
\end{equation*}

\noindent which, in space-time of the charged Hayward black hole with a cosmological constant and surrounded by quintessence and a cloud of
strings, can be written as

\begin{equation}
\begin{aligned}
\mathcal{L}=&+\frac{1}{2}\left(1-a-\frac{2 m r^2}{r^3+2 l^2 m}-\frac{\alpha}{r^{3 \omega _q+1}}+\frac{Q^2}{r^2}-\frac{1}{3}\Lambda  r^2\right)\dot{t}^2
\\
&-\frac{1}{2}\left(1-a-\frac{2 m r^2}{r^3+2 l^2 m}-\frac{\alpha}{r^{3 \omega _q+1}}+\frac{Q^2}{r^2}-\frac{1}{3}\Lambda  r^2\right)^{-1}\dot{r}^2 
\\
&-\frac{r^2}{2}(\dot{\theta}^2 +\sin^2{\theta}\dot{\phi}^2),
\end{aligned}
\label{eq:1.76}
\end{equation}

\noindent where the dot represents the derivative in respect to the proper time $\tau$. The Euler-Lagrange equations are given by

\begin{equation}
\frac{d}{d\tau}\left(\frac{\partial \mathcal{L}}{\partial\dot{x}^\mu}\right)-\frac{\partial \mathcal{L}}{\partial x^\mu}=0.
\label{eq:1.75}
\end{equation}

\noindent For $\mu=0$ and $\mu=3$ in Eq. (\ref{eq:1.75}), with $\mathcal{L}$ given by Eq. (\ref{eq:1.76}), we get, respectively:

\begin{equation}
\dot{t}=\frac{E}{\left(1-a-\frac{2 m r^2}{r^3+2 l^2 m}-\frac{\alpha}{r^{3 \omega _q+1}}+\frac{Q^2}{r^2}-\frac{1}{3}\Lambda  r^2\right)},
\label{eq:1.77}
\end{equation}

\begin{equation}
\dot{\phi}=-\frac{J}{r^2\sin^2\theta},
\label{eq:1.78}
\end{equation}

\noindent where $E$ and $J$ are movement constants, which we can interpret as the energy $E$ and the angular momentum $J$ of the particle that is moving nearby the black hole. 

Restricting our analysis of geodesics to the equatorial plane of the black hole, we reduce Eqs. (\ref{eq:1.77})-(\ref{eq:1.78}) to:

\begin{equation}
\dot{t}=\frac{E}{\left(1-a-\frac{2 m r^2}{r^3+2 l^2 m}-\frac{\alpha}{r^{3 \omega _q+1}}+\frac{Q^2}{r^2}-\frac{1}{3}\Lambda  r^2\right)},
\label{eq:1.79}
\end{equation}

\begin{equation}
\dot{\phi}=-\frac{J}{r^2}.
\label{eq:1.80}
\end{equation}

\noindent Substituting Eqs. (\ref{eq:1.79}) and (\ref{eq:1.80}) into Eq. (\ref{eq:1.76}), we get

\begin{equation}
E^2=\dot{r}^2+V_{eff},
\label{eq:1.82}
\end{equation}

\noindent where

\begin{equation}
V_{eff}=\left(1-a-\frac{2 m r^2}{r^3+2 l^2 m}-\frac{\alpha}{r^{3 \omega _q+1}}+\frac{Q^2}{r^2}-\frac{1}{3}\Lambda  r^2\right)\left(\frac{J^2}{r^2}+L\right),
\label{eq:1.83}
\end{equation}

\noindent This expression gives the effective potential for geodesic motion in a charged Hayward black hole with a cosmological constant, immersed in quintessence and a cloud of strings, with $f(r)$ defined by Eq. (\ref{eq:1.31}) for the general case. Clearly, this general form remains valid for any particular source or combination of sources.

\noindent By means of the relation,

\begin{equation}
\left(\frac{dr}{dt}\right)^2\dot{t}^2=\dot{r}^2,
\label{eq:1.85}
\end{equation}

\noindent into Eq. (\ref{eq:1.82}) and using Eqs. (\ref{eq:1.83}) and Eq. (\ref{eq:1.79}), we get

\begin{equation}
\left(\frac{dr}{dt}\right)^2=f(r)^2\left[1-\frac{f(r)}{E^2}\left(\frac{J^2}{r^2}+L\right)\right].
\label{eq:1.86}
\end{equation}

\subsection{Radial movement of a massless particle}

For the radial movement $(J=0)$ of a massless particle $(L=0)$, Eq. (\ref{eq:1.86}) can be written as
\begin{equation}
\left(\frac{dr}{dt}\right)^2=f(r)^2.
\label{eq:1.87}
\end{equation}

Substituting Eq. (\ref{eq:1.85}) into Eq. (\ref{eq:1.87}), we get the relation between the coordinates $t$ and $r$, which is given by

\begin{equation}
\pm t=\int\frac{1}{1-a-\frac{2 m r^2}{r^3+2 l^2 m}-\frac{\alpha}{r^{3 \omega _q+1}}+\frac{Q^2}{r^2}-\frac{1}{3}\Lambda  r^2}dr.
\label{eq:1.88}
\end{equation}

Using the Eq. (\ref{eq:1.82}), we can obtain the relation between the coordinate $r$ the proper time $\tau$ for the radial movement of a massless particle, which is given by

\begin{equation*}
\left(\frac{dr}{d\tau}\right)^2=E^2,
\end{equation*}
\begin{equation}
\pm\tau=\frac{r}{E}.
\label{eq:1.89}
\end{equation}

\subsection{Radial movement of a massive particle}

Now, let us consider the movement of massive particles  $(L=1)$ in radial trajectories $(J=0)$ nearby the black hole. From Eq. (\ref{eq:1.86}), we obtain
\begin{equation}
\left(\frac{dr}{dt}\right)^2=f(r)^2-\frac{f(r)^3}{E^2}.
\label{eq:1.90}
\end{equation}

Substituting Eq. (\ref{eq:1.87}) into Eq. (\ref{eq:1.90}), we can find the relationship between the coordinates $t$ and  $r$ for the radial movement of the particle:
\begin{equation}
\pm t=\int\frac{dr}{\sqrt{f(r)^2-\frac{f(r)^3}{E^2}}}.
\label{eq:1.91}
\end{equation}

\begin{figure*}[h]
\centering
\begin{minipage}[!]{0.45\linewidth}
\includegraphics[scale=0.6]{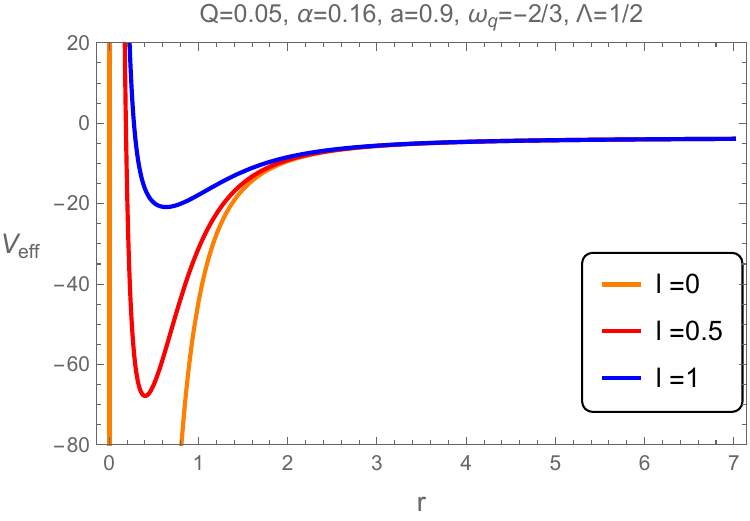}
\vspace{2ex}
      \end{minipage}
\begin{minipage}[!]{0.45\linewidth}
\includegraphics[scale=0.6]{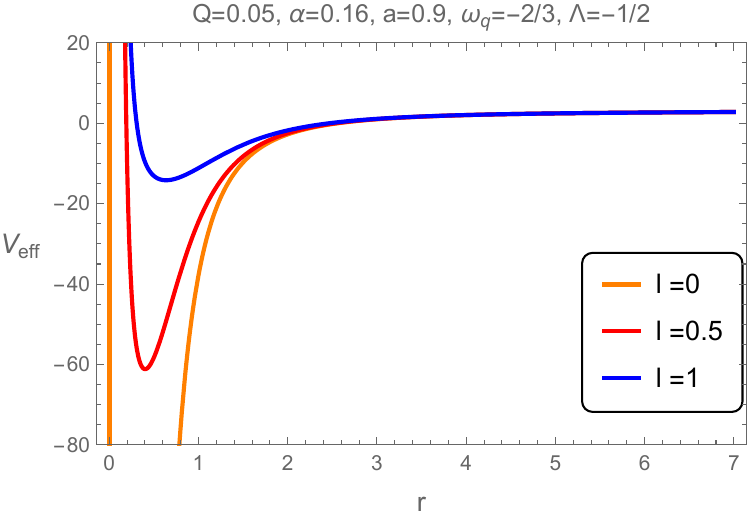}
\vspace{2ex}
     \end{minipage}
\begin{minipage}[!]{0.45\linewidth}
\includegraphics[scale=0.6]{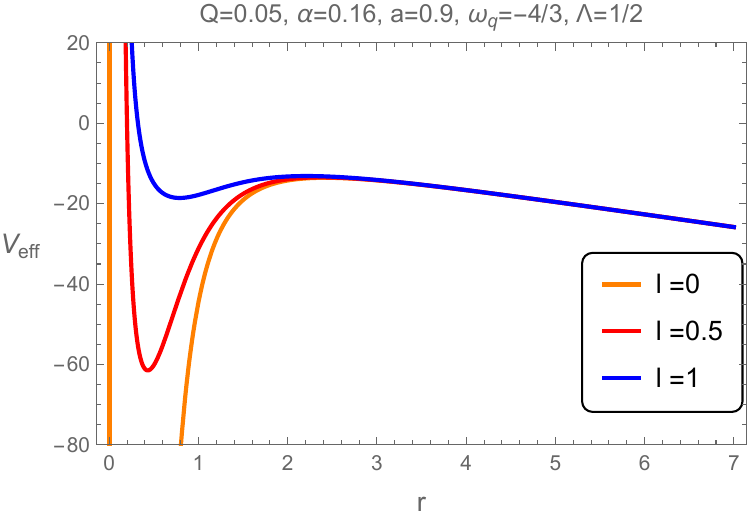}
     \end{minipage}
\begin{minipage}[!]{0.45\linewidth}
\includegraphics[scale=0.6]{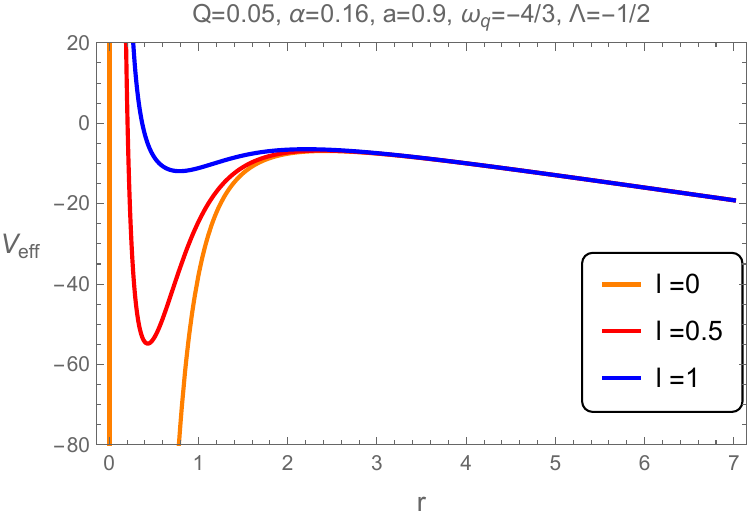}
     \end{minipage}
\caption{Effective potential for non-radial null-like geodesics ($L=0$ and $J^2=20$), for different values of $l$, $\omega_q$ and $\Lambda$.} 
\label{im8}
\end{figure*}


From Eq. (\ref{eq:1.82}), we get the relationship between the proper time $\tau$ and the radial coordinate $r$:
\begin{equation*}
\left(\frac{dr}{d\tau}\right)^2=E^2-f(r),
\end{equation*}
\begin{equation}
\pm\tau=\int\frac{dr}{\sqrt{E^2-f(r)}}.
\label{eq:1.92}
\end{equation}

\subsection{Effective potential}

\begin{figure*}
\centering
\begin{minipage}[!]{0.45\linewidth}
\includegraphics[scale=0.6]{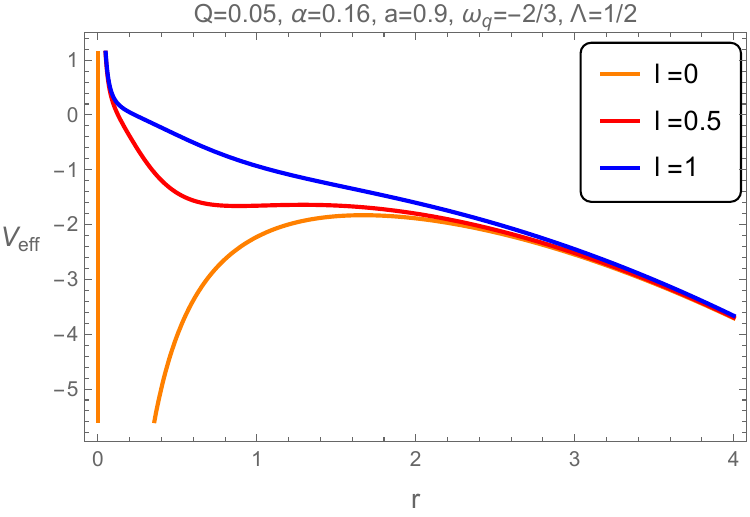}
\vspace{2ex}
      \end{minipage}
\begin{minipage}[!]{0.45\linewidth}
\includegraphics[scale=0.6]{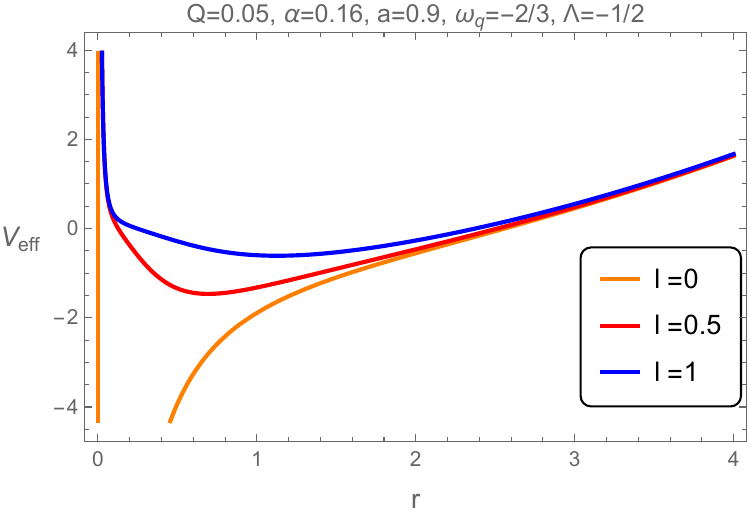}
\vspace{2ex}
     \end{minipage}
\begin{minipage}[!]{0.45\linewidth}
\includegraphics[scale=0.6]{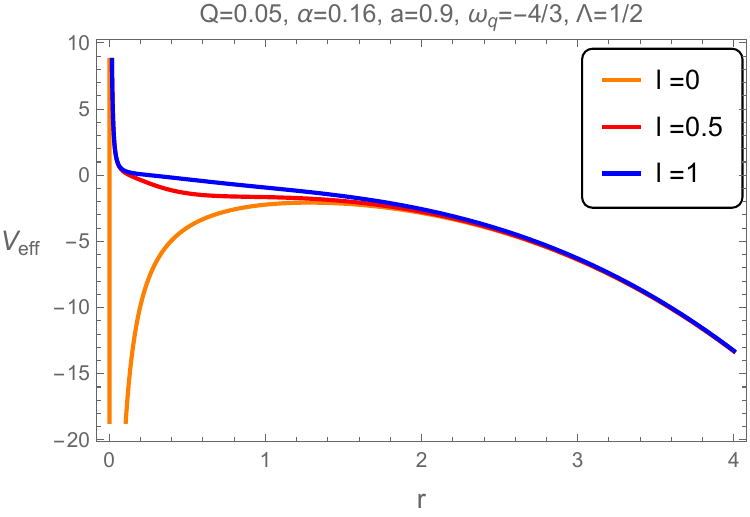}
     \end{minipage}
\begin{minipage}[!]{0.45\linewidth}
\includegraphics[scale=0.6]{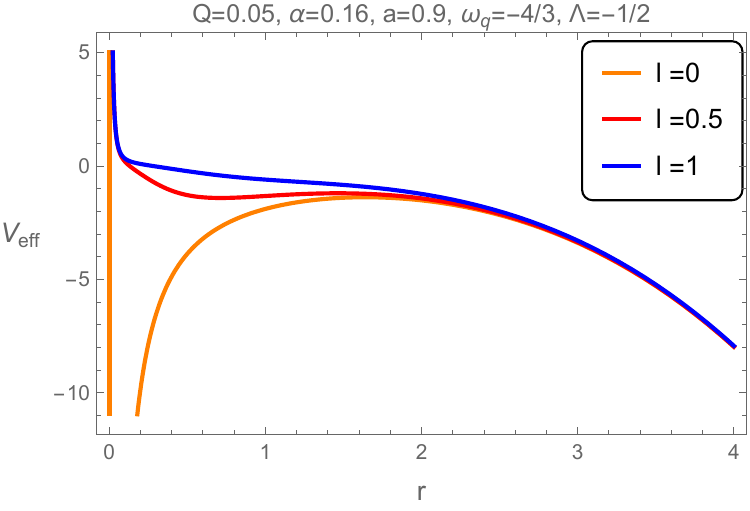}
     \end{minipage}
\caption{Effective potential for radial time-like geodesics ($J^2=0$ and $L=1$), for different values of $l$, $\omega_q$ and $\Lambda$.} 
\label{im9}
\end{figure*}

\begin{figure*}
\centering
\includegraphics[scale=0.9]{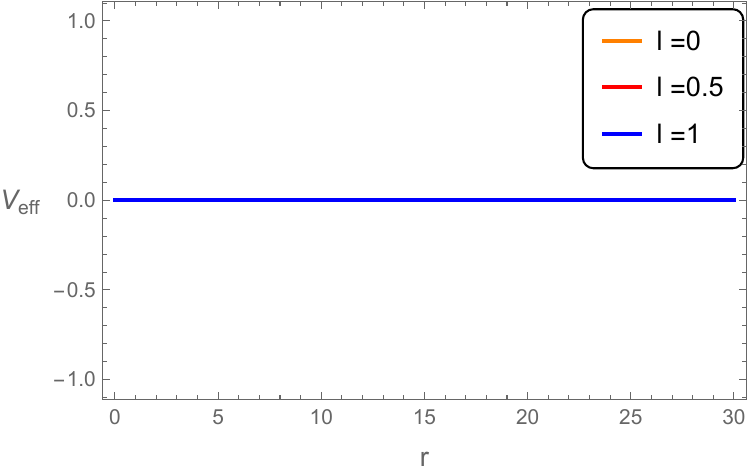}
\caption{Effective potential for radial null-like geodesics  ($J^2=0$ and $L=0$),  for different values of $l$, $\omega_q$ and $\Lambda$.} 
\label{im10}
\end{figure*}

The behavior of the effective potential ($V_{eff}$) of the geodesic motion, given by Eq. (\ref{eq:1.83}), can tell us about the behavior of a massive particle or a massless particle near the black hole. So, in Figs. \ref{im7} to \ref{im10}, we plot the effective potentials for different values of $l$, $\omega_q$ and $\Lambda$ for time-like and null-like geodesics. In some figures, we represent, in detail, $V_{eff}$ near the black hole ($r$ near zero).

In Fig. \ref{im7}, we represent the effective potential for non-radial time-like geodesics ($L=1$ and $J^2=20$), for different values of $l$, $\omega_q$ and $\Lambda$. We can observe that, for $l=0$, there is no stable circular geodesics, since the graphics do not show a local minimum for any value of $Q$, $\omega_q$, $\Lambda$ and $a$. On the other hand, for $l>0$, we can observe the possibility of the existence of stable circular geodesics, depending on the parameter values $Q$, $\omega_q$, $\Lambda$ and $a$. In the region near the black hole, $V_{eff}\rightarrow +\infty$. For regions far from the black hole the effective potential also diverges.

For non-radial null-like geodesics  ($J^2=20$ and $L=0$), Fig. \ref{im8}, we can observe that, in all cases, $V_{eff}\rightarrow +\infty$ for regions near the black hole, $r\rightarrow 0$. For regions far from the black hole, $r\rightarrow \infty$, the $V_{eff}\rightarrow \pm 3.3333$ (graphs in the first row) and $V_{eff}\rightarrow -\infty$ (second line graphs). For $l=0$, there is no stable circular geodesics, since the graphics do not show local minima. The existence of stable circular orbits of massless particle around the black hole depends of $Q$, $\omega_q$, $\Lambda$ and $a$, as can be seen in Fig. \ref{im8}.


In which concerns the behavior of radial time-like geodesics  ($J^2=0$ and $L=1$), Fig. \ref{im9} shows us that the stability of radial
movement does not occur for $l = 0$. On the other hand, for $l>0$, there will always be stable geodesics.

Finally, we can observe in Fig. \ref{im10} that, for radial null-like geodesics ($J^2=0$ and $L=0$), the effective potential is constant and equal to zero.

%
%
\newpage
\section{Concluding remarks}
\label{sec7}

The black hole solutions obtained represent a generalization of the Hayward spacetime and illustrate how each parameter, encoding the source, influences the geometry. From the perspective of Kretschmann scalar regularity, these solutions generally do not exhibit finite curvature at the origin.

In certain instances, Kretschmann scalar analysis indicates regular behavior at the origin for the Hayward spacetime when no extra sources are present, and only in the presence of a cosmological constant. For the Hayward–Kiselev black hole, the spacetime loses regularity at $\omega_q=-2/3$ but remains regular in the phantom energy range $\omega_q=-4/3$. Additionally, adding a cloud of strings or an electric charge disrupts the metric’s regularity, introducing a singularity at the origin.

For non-radial time-like geodesics, stable circular orbits do not exist when $l=0$ for the interval shown in the Fig. \ref{im7}. If 
$l>0$, the presence of such stable circular geodesics depends on the parameters $Q$, $\omega_q$, $\Lambda$ e $a$. In this scenario, the effective potential exhibits divergences both near the black hole and at large distances.

For non‑radial null-like geodesics the effective potential diverges at the origin and approaches a finite value far from the black hole (see Figs. (a) and (b) from Fig. \ref{im8}). When $l=0$ no stable circular geodesics exist because the potential plots lack local minima. Whether massless particles can occupy stable circular orbits depends on the parameters $Q$, $\omega_q$, $\Lambda$ e $a$ (Fig. \ref{im8}). For radial time‑like geodesics, Fig. \ref{im9} indicates that stable orbits always occur when $l>0$. Finally, for radial null-like geodesics the effective potential is constant and vanishes (Fig. \ref{im10}).

Future work may focus on studying the thermodynamics of this black hole, analyzing shadows, and also the analogous solution for the stationary case. Such investigations will further clarify the role of regular black holes in the broader background of general relativity and cosmology.

\section*{Acknowledgements}
V.B. Bezerra is partially
supported by CNPq-Brazil ( Conselho Nacional de Desenvolvimento Científico e Tecnológico) through Research Project No. 307211/2020-7.

F. F. Nascimento and J. M. Toledo acknowledge Departamento de Fisica, Universidade Federal da Paraiba, for hospitality.

\appendix

\bibliographystyle{elsarticle-num} \biboptions{numbers}
\bibliography{refs}

\begin{thebibliography}{10}
\expandafter\ifx\csname url\endcsname\relax
  \def\url#1{\texttt{#1}}\fi
\expandafter\ifx\csname urlprefix\endcsname\relax\def\urlprefix{URL }\fi
\expandafter\ifx\csname href\endcsname\relax
  \def\href#1#2{#2} \def\path#1{#1}\fi

\bibitem{schwarzschild1916uber}
K.~Schwarzschild, Uber das gravitationsfeld eines massenpunktes nach der
  einstein'schen theorie, Berlin. Sitzungsberichte 18.

\bibitem{reissner1916eigengravitation}
H.~Reissner, {\"U}ber die eigengravitation des elektrischen feldes nach der
  einsteinschen theorie, Annalen der Physik 355~(9) (1916) 106--120.

\bibitem{nordstrom1918een}
G.~Nordstr{\"o}m, Een en ander over de energie van het zwaartekrachtsveld
  volgens de theorie van einstein, Koninklijke Akademie van Wetenschappen te
  Amsterdam, 1918.

\bibitem{kerr1963gravitational}
R.~P. Kerr, Gravitational field of a spinning mass as an example of
  algebraically special metrics, Physical review letters 11~(5) (1963) 237.

\bibitem{newman1965metric}
E.~T. Newman, E.~Couch, K.~Chinnapared, A.~Exton, A.~Prakash, R.~Torrence,
  Metric of a rotating, charged mass, Journal of mathematical physics 6~(6)
  (1965) 918--919.

\bibitem{neves2017relatividade}
J.~Neves, Relatividade bem comportada: buracos negros regulares, Revista
  Brasileira de Ensino de F{\'\i}sica 39.

\bibitem{ayon2000bardeen}
E.~Ay{\'o}n-Beato, A.~Garc{\i}a, The bardeen model as a nonlinear magnetic
  monopole, Physics Letters B 493~(1-2) (2000) 149--152.

\bibitem{bardeen1968non}
J.~M. Bardeen, Non-singular general-relativistic gravitational collapse, in:
  Proc. Int. Conf. GR5, Tbilisi, Vol. 174, 1968, p. 174.

\bibitem{hayward2006formation}
S.~A. Hayward, Formation and evaporation of nonsingular black holes, Physical
  review letters 96~(3) (2006) 031103.

\bibitem{frolov2016notes}
V.~P. Frolov, Notes on nonsingular models of black holes, Physical Review D
  94~(10) (2016) 104056.

\bibitem{kiselev2003quintessence}
V.~V. Kiselev, Quintessence and black holes, Classical and Quantum Gravity
  20~(6) (2003) 1187.

\bibitem{riess1998observational}
A.~G. Riess, A.~V. Filippenko, P.~Challis, A.~Clocchiatti, A.~Diercks, P.~M.
  Garnavich, R.~L. Gilliland, C.~J. Hogan, S.~Jha, R.~P. Kirshner, et~al.,
  Observational evidence from supernovae for an accelerating universe and a
  cosmological constant, The astronomical journal 116~(3) (1998) 1009.

\bibitem{riess1999bvri}
A.~G. Riess, R.~P. Kirshner, B.~P. Schmidt, S.~Jha, P.~Challis, P.~M.
  Garnavich, A.~A. Esin, C.~Carpenter, R.~Grashius, R.~E. Schild, et~al., Bvri
  light curves for 22 type ia supernovae, The Astronomical Journal 117~(2)
  (1999) 707.

\bibitem{perlmutter1999measurements}
S.~Perlmutter, G.~Aldering, G.~Goldhaber, R.~A. Knop, P.~Nugent, P.~G. Castro,
  S.~Deustua, S.~Fabbro, A.~Goobar, D.~E. Groom, et~al., Measurements of
  $\omega$ and $\lambda$ from 42 high-redshift supernovae, The Astrophysical
  Journal 517~(2) (1999) 565.

\bibitem{ade2014planck}
P.~A. Ade, N.~Aghanim, M.~Alves, C.~Armitage-Caplan, M.~Arnaud, M.~Ashdown,
  F.~Atrio-Barandela, J.~Aumont, H.~Aussel, C.~Baccigalupi, et~al., Planck 2013
  results. i. overview of products and scientific results, Astronomy \&
  Astrophysics 571 (2014) A1.

\bibitem{letelier1979clouds}
P.~S. Letelier, Clouds of strings in general relativity, Physical Review D
  20~(6) (1979) 1294.

\bibitem{ghosh2015nonsingular}
S.~G. Ghosh, A nonsingular rotating black hole, The European Physical Journal C
  75~(11) (2015) 532.

\bibitem{Ling_2023}
Y.~Ling, M.-H. Wu, \href{https://doi.org/10.1088/1361-6382/acc0c9}{Regular
  black holes with sub-planckian curvature}, Classical and Quantum Gravity
  40~(7) (2023) 075009.
\newblock \href {http://dx.doi.org/10.1088/1361-6382/acc0c9}
  {\path{doi:10.1088/1361-6382/acc0c9}}.
\newline\urlprefix\url{https://doi.org/10.1088/1361-6382/acc0c9}

\bibitem{rodrigues2022bardeenclounds}
M.~E. Rodrigues, H.~A. Vieira, Bardeen solution with a cloud of strings,
  Physical Review D 106~(8) (2022) 084015.

\bibitem{rodrigues2018bardeen}
M.~E. Rodrigues, M.~V. d.~S. Silva, Bardeen regular black hole with an electric
  source, Journal of Cosmology and Astroparticle Physics 2018~(06) (2018) 025.

\bibitem{santos2024regular}
L.~C. Santos, Regular black holes from kiselev anisotropic fluid, The European
  Physical Journal C 84~(12) (2024) 1318.

\bibitem{nascimento2024somebardeen}
F.~Nascimento, P.~H. Morais, J.~Toledo, V.~Bezerra, Some remarks on bardeen-ads
  black hole surrounded by a fluid of strings, General Relativity and
  Gravitation 56~(7) (2024) 86.

\bibitem{nascimento2023some}
F.~Nascimento, V.~Bezerra, J.~Toledo, Some remarks on hayward black hole
  surrounded by a cloud of strings, Annals of Physics (2023) 169548.

\bibitem{lemos2011regular}
J.~P. Lemos, V.~T. Zanchin, Regular black holes: Electrically charged
  solutions, reissner-nordstr{\"o}m outside a de sitter core, Physical Review
  D—Particles, Fields, Gravitation, and Cosmology 83~(12) (2011) 124005.

\bibitem{nascimento2024black}
F.~F.~d. Nascimento, V.~B. Bezerra, J.~d.~M. Toledo, Black holes with a cloud
  of strings and quintessence in a non-linear electrodynamics scenario,
  Universe 10~(11) (2024) 430.

\bibitem{bronnikov2001regular}
K.~A. Bronnikov, Regular magnetic black holes and monopoles from nonlinear
  electrodynamics, Physical Review D 63~(4) (2001) 044005.

\bibitem{molina2021thermodynamics}
M.~Molina, J.~Villanueva, On the thermodynamics of the hayward black hole,
  Classical and Quantum Gravity 38~(10) (2021) 105002.

\bibitem{fan2016construction}
Z.-Y. Fan, X.~Wang, Construction of regular black holes in general relativity,
  Physical Review D 94~(12) (2016) 124027.

\bibitem{bronnikov2017comment}
K.~A. Bronnikov, Comment on “construction of regular black holes in general
  relativity”, Physical Review D 96~(12) (2017) 128501.

\bibitem{toshmatov2018comment}
B.~Toshmatov, Z.~Stuchl{\'\i}k, B.~Ahmedov, Comment on “construction of
  regular black holes in general relativity”, Physical Review D 98~(2) (2018)
  028501.

\bibitem{rodrigues2022bardeen}
M.~E. Rodrigues, M.~V. d.~S. Silva, H.~A. Vieira, Bardeen-kiselev black hole
  with a cosmological constant, Physical Review D 105~(8) (2022) 084043.

\bibitem{d2022introducing}
R.~d'Inverno, J.~Vickers, Introducing Einstein's Relativity: A Deeper
  Understanding, Oxford University Press, 2022.

\end{thebibliography}






\end{document}